# Asymptotic scaling from strong coupling in 2-d lattice chiral models.

Massimo Campostrini, Paolo Rossi, and Ettore Vicari [a] [*]

[a]Dipartimento di Fisica dell'Università di Pisa, Italy.

Two dimensional $N = \infty$ lattice chiral models are investigate by a strong coupling analysis. Strong coupling expansion turns out to be predictive for the evaluation of continuum physical quantities, to the point of showing asymptotic scaling (within 5%).

## 1. Introduction

The study of 2-d principal chiral models,

$$S = \int d^2x \, \frac{1}{T} \, \text{Tr} \, \partial_\mu U(x) \partial_\mu U^\dagger(x) \qquad (1)$$

($U \in SU(N)$), is strongly motivated by the analogies with 4-d non-Abelian gauge theories, e.g. asymptotic freedom and the large-$N$ limit represented by a sum over planar graphs. According to the conjectured S-matrix [1] and large-N factorization, these theories should describe free particles in the limit $N \to \infty$. However analyzing the Green's functions of the theory, the realization of such physical properties appears not trivial at all. Indeed in the large-N limit the following scenario emerges [2]: The Lagrangian fields $U$, playing the role of non-interacting hadrons, are constituted by two confined particles, which appear free in the large momentum limit, due to asymptotic freedom.

Recent numerical studies of 2-d lattice $SU(N)$ chiral models with nearest-neighbor interaction

$$S_L = -2N\beta \sum_{x,\mu} \text{Re} \, \text{Tr} \left[ U(x) \, U^\dagger(x+\mu) \right] \qquad (2)$$

($\beta = 1/NT$), have shown the existence of a scaling region, where continuum predictions for dimensionless ratios of physical quantities are verified [3,4,2]. The scaling region begins at small values of the correlation length, well within the expected region of convergence of strong-coupling expansion. Furthermore, in the whole scaling region the fundamental mass agrees, within few per cent, with the asymptotic scaling predictions in the energy scheme [2].

As a matter of fact, this may be considered as an evidence for asymptotic scaling within the strong-coupling regime, motivating a test of scaling and asymptotic scaling by strong coupling computations. As a byproduct, strong-coupling series can be analyzed to investigate the critical behavior of the $N = \infty$ lattice theory.

We generated large-$N$ strong coupling series of the free energy up to $O\left(\beta^{18}\right)$, and of the fundamental Green's function

$$G(x) = \langle \frac{1}{N} \text{Re} \, \text{Tr} \left[ U(x)U(0)^\dagger \right] \rangle \qquad (3)$$

up to $O\left(\beta^{15}\right)$, from which we extracted several physical quantities, e.g. the fundamental mass and the second moment [5].

## 2. The large-$N$ phase transition.

Lattice chiral models have a peak in the specific heat that becomes more and more pronounced with increasing $N$ [2]. Fig. 1 shows Monte Carlo data of the specific heat of some large-$N$ $SU(N)$ and $U(N)$ models. We recall that $U(N)$ and $SU(N)$ models should have the same large-$N$ limit. The behavior of the specific heat in $U(N)$ and $SU(N)$ models at large $N$ should be an indication of a phase transition at $N = \infty$. With increasing $N$, the positions of the peaks $\beta_{peak}$ in SU($N$) and U($N$) converge from opposite directions. An estimate of the critical coupling $\beta_c$ was obtained by extrapolating $\beta_{peak}(N)$ to $N \to \infty$ using a finite $N$ scaling Ansatz [6]

$$\beta_{peak}(N) \simeq \beta_c + cN^{-\epsilon}, \qquad (4)$$

[*]This work was partially supported by MURST, and E.C. contract CHRX-CT92-0051.

which resembles a more rigorouos finite size scaling relationship. The above Ansatz was suggested by the idea that the parameter $N$ may play a role quite analogous to the volume in the ordinary systems close to the criticality. The extrapolation according to (4) of our $U(N)$ and $SU(N)$ data (at $N = 9, 15, 21$ for $U(N)$ and $N = 9, 15, 21, 30$ for $SU(N)$) gave $\beta_c = 0.3057(3)$. Notice that at $N = \infty$ the value of the correlation length in the fundamental channel at $\beta_c$ is finite $\xi^{(c)} \simeq 2.8$.

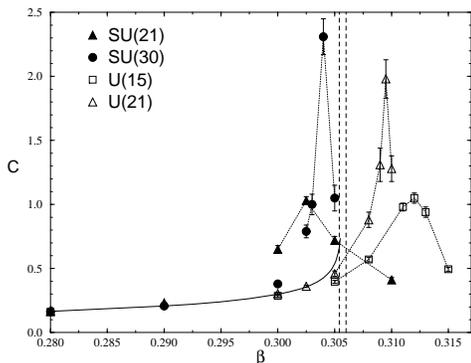

Figure 1. Specific heat vs. $\beta$. The estimate of the $N = \infty$ critical $\beta$ is indicated by the vertical dashed lines.

The existence of this phase transition is confirmed by an analysis of the $N = \infty$ 18th order strong-coupling series of $C$, based on the method of the integral approximants [7]. Such an analysis gave quite stable results showing a second order critical behavior:

$$C \sim |\beta - \beta_c|^{-\alpha}, \qquad (5)$$

with $\beta_c = 0.3058(3)$ and $\alpha = 0.23(3)$, in agreement with the extrapolation of Monte Carlo data. Fig. 1 shows that simulation data of $C$ approach, for growing $N$, the curve (represented by the full line) obtained from the resummation of the $N = \infty$ strong-coupling series.

At finite $N$ $U(N)$ models should experience a phase transition, driven by the $U(1)$ degrees of freedom corresponding to the determinant of $U(x)$, with a pattern similar to the XY model [8]. The location of this phase transition is beyond the specific heat peak [6] and may eventually converge to $\beta_c$ in the large-$N$ limit, as argued by Green and Samuel [8]. According to this conjecture the large-$N$ limit would change the order of the determinant phase transition from an infinite order of the Kousterlitz Thouless mechanism to a second order with divergent specific heat.

## 3. Scaling and asymptotic scaling from strong coupling.

In spite of the existence of a phase transition at $N = \infty$, large-$N$ Monte Carlo data show scaling and asymptotic scaling (in the energy scheme) even for $\beta$ smaller then the peak of the specific heat, suggesting an effective decoupling of the modes responsible for the large-$N$ phase transition from those determining the physical continuum limit. This fact motivates a test of scaling and asymptotic scaling at $N = \infty$ based only on strong coupling computations, given that strong coupling expansion should converge for $\beta < \beta_c$.

In Fig. 2 we plot the dimensionless ratio $M/M_G$ vs. the correlation length $\xi_G \equiv 1/M_G$, where $M$ is the fundamental mass and $\xi_G$ is the correlation length defined from the second moment of the fundamental Green's function, as obtained from our $N = \infty$ strong-coupling series [5]. The ratio $M/M_G$ is quite stable for a large region of values of $\xi_G$ and in good agreement (well within 1%) with the continuum large-$N$ value extrapolated by Monte Carlo data $M/M_G = 0.991(1)$ [2].

In order to test asymptotic scaling we change variable from $\beta$ to $\beta_E = (8E)^{-1}$ [9], evaluating the internal energy $E$ from its strong-coupling series. The two loop renormalization group and a Bethe Ansatz evaluation of the mass $\Lambda$-parameter ratio [10] lead to the following large-$N$ asymptotic scaling prediction in the $\beta_E$ scheme:

$$M \simeq 16\sqrt{\frac{\pi}{e}} \exp\left(\frac{\pi}{4}\right) \Lambda_{E,2l}(\beta_E),$$
$$\Lambda_{E,2l}(\beta_E) = \sqrt{8\pi\beta_E}\exp(-8\pi\beta_E),$$
$$\beta_E = \frac{1}{8E}. \qquad (6)$$

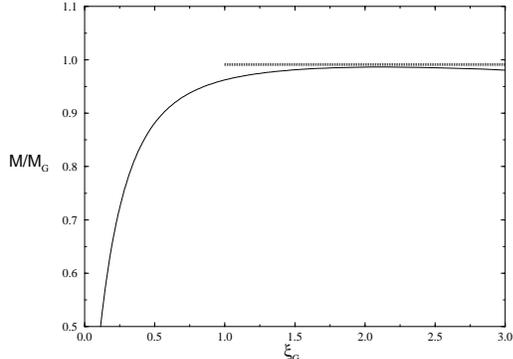

Figure 2. $M/M_G$ vs. $\xi_G \equiv 1/M_G$. The dashed line represents the continuum result from Monte Carlo data.

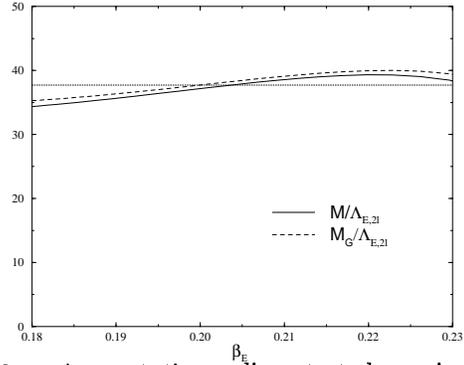

Figure 3. Asymptotic scaling test by using strong-coupling estimates. The dotted line represents the exact result (6).

In Fig. 3 the strong-coupling estimates of $M/\Lambda_{E,2l}$ and $M_G/\Lambda_{E,2l}$ are plotted vs. $\beta_E$, for a region of coupling corresponding to correlation lengths from about one to three. These quantities agree with the exact continuum prediction within about 5% in the whole region.

Notice that the good behavior of the large-$N$ $\beta$-function in the $\beta_E$ scheme, and therefore the fact that physical quantities appear to be well behaved functions of the energy, together with the critical behavior (5) lead to a non-analytical zero at $\beta_c$ of the $\beta$-function in the standard scheme:

$$\beta_L(T) \sim |\beta - \beta_c|^\alpha . \qquad (7)$$

This was also confirmed by an analysis of the strong-coupling series of the magnetic susceptibility $\chi$ and $M_G^2$, which supported the following relations

$$\frac{d\ln\chi}{d\beta} \sim \frac{d\ln M_G^2}{d\beta} \sim |\beta - \beta_c|^{-\alpha} \qquad (8)$$

in the neighbourhood of $\beta_c$.

We would like to mention that similar results were obtained for 2-d chiral models on the honeycomb lattice, which is expected to belong to the same universality class with respect to the continuum limit. In particular asymptotic scaling was verified within about 10% by the correspoding strong coupling expansion.

## REFERENCES

1. E. Abdalla, M.C.B. Abdalla and A. Lima-Santos, Phys. Lett. **140B**, 71 (1984); P. Wiegmann, Phys. Lett. **141B**, 217 (1984).
2. P. Rossi and E. Vicari, Phys. Rev. **D**, 49 (1994) 6072; **D** 50 (1994) 4718 (E).
3. P. Rossi and E. Vicari, Phys. Rev. **D**, 49 (1994) 1621.
4. R.R. Horgan and I.T.Drummond, Nucl. Phys. B (Proc. Suppl.) 34 (1994) 686.
5. M. Campostrini, P. Rossi and E. Vicari, "Asymptotic scaling from strong coupling", IFUP-TH 36/94 (1994).
6. M. Campostrini, P. Rossi and E. Vicari, "Large-N phase transition in lattice 2-d principal chiral models", IFUP-TH 56/94.
7. D. L. Hunter and G. A. Baker Jr., Phys. Rev. **B 49** (1979) 3808; M. E. Fisher and H. Au-Yang, J. Phys. **A12** (1979) 1677.
8. F. Green and S. Samuel, Phys. Lett. **103B**, 110 (1981).
9. G. Parisi, Proceedings of the XXth Conference on High Energy Physics, Madison, Wisconsin, 1980.
10. J. Balog, S. Naik, F. Niedermayer, and P. Weisz, Phys. Rev. Lett. **69**, 873 (1992).